\documentclass[20pt]{article}
\usepackage[utf8]{inputenc}
\usepackage{amsmath}
\usepackage{geometry}
\usepackage{setspace}
\usepackage{cite}
\usepackage{authblk} 
\usepackage{cite}
\usepackage{setspace}
\usepackage[utf8]{inputenc}
\usepackage[T1]{fontenc}
\usepackage{times}  
\usepackage{textcomp}  
\DeclareUnicodeCharacter{2212}{\textminus}
\usepackage{upgreek,float}
\usepackage{amsmath}  
\usepackage{graphicx} 
\usepackage{float}      
\usepackage{caption}
\usepackage{subcaption} 
\usepackage{indentfirst} 
\usepackage{authblk}

\providecommand{\keywords}[1]
{
  \small	
  \textbf{\textit{Keywords---}} #1
}

\title{Effect of near-earth thunderstorm electric field on the flux of cosmic ray air showers in LHAASO-KM2A}
\author{
  C. Yang\textsuperscript{1}, 
  X. X. Zhou\textsuperscript{1,}\footnote{Corresponding author: zhouxx@swjtu.edu.cn} 
     , 
  H. H. He\textsuperscript{2,3,}\footnote{Corresponding author: hhh@ihep.ac.cn}   ,
  D. H. Huang\textsuperscript{1}, 
  X. J. Chen\textsuperscript{1}, 
  T. Zhou\textsuperscript{1}, 
  K. J. Guo\textsuperscript{1}
}

\affil{\textsuperscript{1}{\itshape\footnotesize School of Physical Science and Technology, Southwest Jiaotong University, Chengdu 610031, China}\\
 \textsuperscript{2}{\itshape\footnotesize Key Laboratory of Particle Astrophysics, Institute of High Energy Physics, Chinese Academy of Sciences, Beijing 100049, China}\\
 \textsuperscript{3}{\itshape\footnotesize University of Chinese Academy of Sciences, Beijing 100049, China}}

\date{}

\begin{document}

\renewenvironment{abstract}
 {\normalsize 
  \begin{center} 
  \Large  \bfseries \abstractname 
  \end{center}
  \quotation}
 {\endquotation}

\captionsetup[figure]{labelfont={bf},name={Fig.},labelsep=space}
\maketitle
\large
\begin{abstract}

 The Large High Altitude Air Shower Observatory (LHAASO) is located at Haizi Mountain (4410 m a. s. l.), Daocheng, Sichuan province, China. Due to its high-altitude location with frequent thunderstorm activities, the LHAASO is suited for studying the effects of near-earth thunderstorm electric fields on cosmic ray air showers. In this paper, Monte Carlo simulations are performed with CORSIKA and G4KM2A to analyze the flux variations of cosmic ray air showers detected by the kilometer-square array of LHAASO (LHAASO-KM2A) during thunderstorms. The strength, polarity, and layer thickness of atmospheric electric field (AEF) during thunderstorm are found to be associated with the shower rate variations. The flux of shower events satisfying trigger conditions of the LHAASO-KM2A increases with field intensity, particularly within negative fields, and the enhanced amplitude is more than 5\% in -600 V/cm and 12\% in AEFs of -1000 V/cm, whereas it increases by only 1\% and 7\% in equivalent positive fields, respectively. While in positive fields ranging from 0 to 400 V/cm, the shower rate decreases with smaller amplitudes. Furthermore, the shower rate increases dramatically with the AEF layer thickness until a certain value, above which the variation trend slows down. The dependence of the trigger rate variation on the primary zenith angle has also been revealed, increasing in lower zenith angle ranges and showing opposite behaviors in higher ones. Additionally, we study that the relationship between the trigger rate variations and the primary energies, and find the enhanced amplitude of the shower rate decreases with increasing primary energy. Simultaneously, the shower events with lower primary energy show a significant increase, whereas events with higher primary energy are hardly affected during thunderstorms. These simulation results offer valuable insights into the variation of the trigger rate detected by LHAASO-KM2A during thunderstorms and are beneficial for understanding the acceleration mechanisms of secondary charged particles caused by AEF.\\
\keywords{Near-earth thunderstorm electric fields, Cosmic rays, Extensive air showers, LHAASO-KM2A}
\end{abstract}

\section{Introduction}

Thunderstorms are common weather events at high-altitude areas, accompanied by lightning activities that likely pose a threat to human life and property. Thunderclouds distribute roughly within the altitude scope of 4-12 km in the atmosphere. The charge structure in thunderclouds is complex, and the electric field strength can be up to 1000 V/cm, or even higher \cite{ref1, ref2, ref3}. It was first suggested by Wilson \cite{ref4} in 1924 that strong electric fields within thunderclouds could accelerate electrons, which have tiny mass, and proposed the concept of "runaway electrons". Scientists have made several attempts over the following decades to find the runaway electrons predicted during thunderstorms. It was not until 1992 that a significant breakthrough was achieved by Gurevich et al. \cite{ref5}, who found that cosmic ray secondary particles accelerated by the strong AEF gain energy exceeding that lost in ionization and bremsstrahlung. Meanwhile, these electrons are capable of generating new electrons continuously through the ionization of the air molecules, eventually giving rise to an avalanche. This process is known as the theory of runaway breakdown (RB), now commonly referred to as relativistic runaway electron avalanche (RREA) \cite{ref6,ref7}.

According to the proposed theory by Dwyer \cite{ref8} and Symbalisty et al. \cite{ref9}, the field strength threshold ($E_{\text{th}}$) required to trigger the RREA process decreases with increasing altitude and can be expressed by \(E_{\text{th}} = E_0 e^{-z/z_0}\), where $z$ is the height above sea level (in km), $z_0$ (~8.4 km) is the scale height. The threshold at sea level is about 2800 V/cm and it decreases to $\approx$ 1660 V/cm at $z$ = 4.41 km (the height of LHAASO), which means that a very large AEF strength is needed to trigger the RREA process.

When a primary cosmic ray enters the atmosphere, it will generate a large number of secondary particles via hadronic and electromagnetic cascades, which are distributed over a range many kilometers wide. This phenomenon is referred to as extensive air shower (EAS) \cite{ref10}. Due to acceleration or deceleration caused by strong AEF, the number and energy of secondary particles in EAS could be modified when they cross several-kilometer thundercloud layers. Simultaneously, the space-time distribution of charged particles at the ground level will also be changed because of the deflection effect of AEF. For ground-based experiments, the measurement and traceability of cosmic rays will certainly be affected \cite{ref11}.

Nowadays, the high-energy phenomenon originating in the terrestrial atmosphere during thunderstorms has become a hot topic in the interdisciplinary domain of atmospheric physics and cosmic ray physics. To date, some space-borne experiments, such as BATSE \cite{ref12,ref13}, AGILE \cite{ref14,ref15}, ASIM \cite{ref16,ref17}, Fermi \cite{ref18,ref19,ref20}, RELEC \cite{ref21}, and Insight-HXMT \cite{ref22}, have detected thousands of terrestrial gamma-ray flashes (TGFs), sub-millisecond gamma-ray emissions originated from bremsstrahlung by runaway electrons, which can be explained by the RREA mechanism \cite{ref23}. Furthermore, scientists have carried out a large number of ground-based experiments at high altitude to study high-energy particles in the atmosphere. In the Baksan Carpet array \cite{ref24}, EAS-TOP \cite{ref25}, AS$\gamma$ \cite{ref26}, ASEC \cite{ref27,ref28,ref29}, and detectors on Mount Fuji \cite{ref30}, sudden enhancements of cosmic ray secondary particles during thunderstorms, known as thunderstorm ground enhancements (TGEs), have been detected. Their results indicate that the intensity enhancements of ground cosmic rays are associated with the electric fields in thunderclouds and the RREA process is responsible for huge TGEs.

By analyzing the data from the ARGO-YBJ experiment in scaler mode, the counting rates of showers with different particle multiplicities have been found to be strongly dependent on the intensity and polarity of the electric field\cite{ref31}. Additionally, an unusual phenomenon was discovered where the flux of cosmic ray secondary particles decreases rather than increases in a certain range of positive fields, which is due to the asymmetry in the number and energy of electrons and positrons in EAS \cite{ref32}.

In general, the trigger conditions of air shower arrays are determined by the energy and space-time distribution of shower particles. During thunderstorms, these particles could be significantly influenced, resulting in variations of shower rates that meet the trigger conditions in ground-based experiments. By analyzing the data recorded by the ARGO-YBJ detector during thunderstorms, Axikegu et al. \cite{ref11} found that flux variations in detected shower events are correlated to the intensity and polarity of the AEF and also strongly dependent on the primary zenith angle. Aharonian et al. \cite{ref33} reported flux variations of cosmic ray air showers increased during thunderstorm 20210610 and revealed the average shower rate increase for smaller zenith angle ranges and decrease for larger ones. These findings highlight the complexity and diversity of observational results of cosmic ray air shower variations during thunderstorms.

To deeply interpret the observed results and understand the underlying mechanisms in experiments, Zhou et al. \cite{ref32} simulated the development of the electrons and positrons in EAS during thunderstorms and indicated that the number of particles depends on the electric field intensity and polarity. In addition, there have been some reports on the energy and lateral distribution of secondary particles during thunderstorms. Chilingarian et al. \cite{ref34} analyzed the energy spectra of gamma rays in the presence of AEF. They found that the power law shape of the gamma-ray differential energy spectra tends to soften with increasing electric field strength. Yan et al. \cite{ref35} investigated the field effects on the energy of ground cosmic rays and found the energy distribution of electrons changes in the field. In the low energy region, the total number of electrons and positrons increases significantly, while at high energies, it does not change obviously. The field effects on the lateral density of secondary positrons and electrons were analyzed by  K. G. Axi et al. \cite{ref36}, and they found that the lateral distribution becomes wider during thunderstorms. For more simulation results of the relationship between cosmic rays and thunderstorm electric fields, please refer to other papers \cite{ref8,ref37,ref38,ref39}.

Up to now, considerable advances have been achieved in experimental and theoretical research on the intensity variation of ground cosmic rays during thunderstorms. Due to the complex charge structure in thunderclouds, there is no clear and comprehensive picture of cosmic ray variations during thunderstorms. For example, how do cosmic ray intensities vary during different thunderstorm events? Within the same thunderstorm, how do changes in cosmic rays occur with different energies or directions? The physical mechanisms of acceleration and deflection of cosmic ray secondary charged particles by thunderstorm electric fields remain unanswered. These questions are still far from being fully understood. In order to learn more about the AEF effect on cosmic rays, more experimental observations and simulation studies are needed.

LHAASO has distinctive advantages for studying cosmic ray variations during thunderstorms, owing to its location at a high-altitude area with frequent thunderstorms. In this paper, the CORSIKA code is used to simulate the EAS development, and the G4KM2A code is employed to simulate the response of the LHAASO-KM2A detector. By using these simulation codes, the influences of thunderstorm electric fields on the shower rates are studied in detail.

\section{Simulation parameters}
To simulate the development of EAS initiated by high-energy cosmic rays, CORSIKA has been used with version 7.7410 \cite{ref40}. The QGSJETII-04 model is adopted to simulate hadronic interactions for energies higher than 80 GeV, while the GHEISHA code is used for lower energy interactions. Proton primaries are assumed with arrival directions evenly distributed across the sky, with zenith angles ranging from 0° to 60° and azimuth angles from 0° to 360°. Based on the observational energy threshold of the LHAASO-KM2A in shower mode \cite{ref41,ref42}, primary protons are simulated over a wide energy range from 1 to 10$^5$ TeV, following a power-law spectrum with an index of -2.7. Considering the effects of electric fields on acceleration, the energy cutoff for secondary electrons, positrons, and photons is set to the lowest threshold allowed by CORSIKA, which is 50 keV. This implies that secondary particles will be tracked until their energies reduce to this threshold. Furthermore, horizontal and vertical intensity components of the magnetic field are set to $B$$_x$ = 34.8 $\mu$T and $B$$_z$ = 36.2 $\mu$T, sourced from NOAA's National Centers for Environmental Information (NCEI) \cite{ref43}.

Since secondary charged particles quickly dissipate energy through radiation and ionization, the effects of electric fields can be neglected on charged particles measured at ground level when thunderclouds are far from the detector \cite{ref44}. Therefore, in this work, only the near-earth electric field is considered. Additionally, the intensities of electric fields within most thunderstorms are well below the relativistic runaway electron avalanche (RREA) threshold. A series of electric field values ranging from ±1200 V/cm are chosen for simulation, uniformly distributed vertically within an atmospheric layer extending from the LHAASO level (4410 m) up to 5410 m. Here, positive electric fields are defined as those that accelerate positrons downward towards the Earth.

The LHAASO experiment is located at an altitude of 4410 m above sea level (29° 21' 27.56" N, 100° 08' 19.66" E) in Daocheng, China, where the measured atmospheric depth is 598.0 g/cm$^2$. It consists of three detector arrays: a kilometer square array (KM2A), a water Cherenkov detector array (WCDA), and a wide-field-of-view Cherenkov/fluorescence telescope array (WFCTA). The KM2A (as shown in Fig. 1) comprises 5216 electromagnetic particle detectors (EDs) and 1188 muon detectors (MDs). The ED array is divided into two parts: the central array consists of 4911 EDs spaced 15 meters apart in a circular area, and the guarding ring array forms the outskirts with 305 EDs spaced 30 meters apart \cite{ref41,ref45}. As of July 2021, the full KM2A array has been installed and is operating stably.
The trigger mode for the full KM2A array requires at least 20 EDs or MDs fired within a 400 ns time window. The data acquisition system (DAQ) records 5000 \text{ns} of data before and after the trigger time from all EDs and MDs. The ED signals are used to reconstruct parameters of the shower such as the core position, arrival direction, and primary energy \cite{ref41,ref46}.\par

\begin{figure}[H] 
    \centering
    \includegraphics[width=0.7\textwidth]{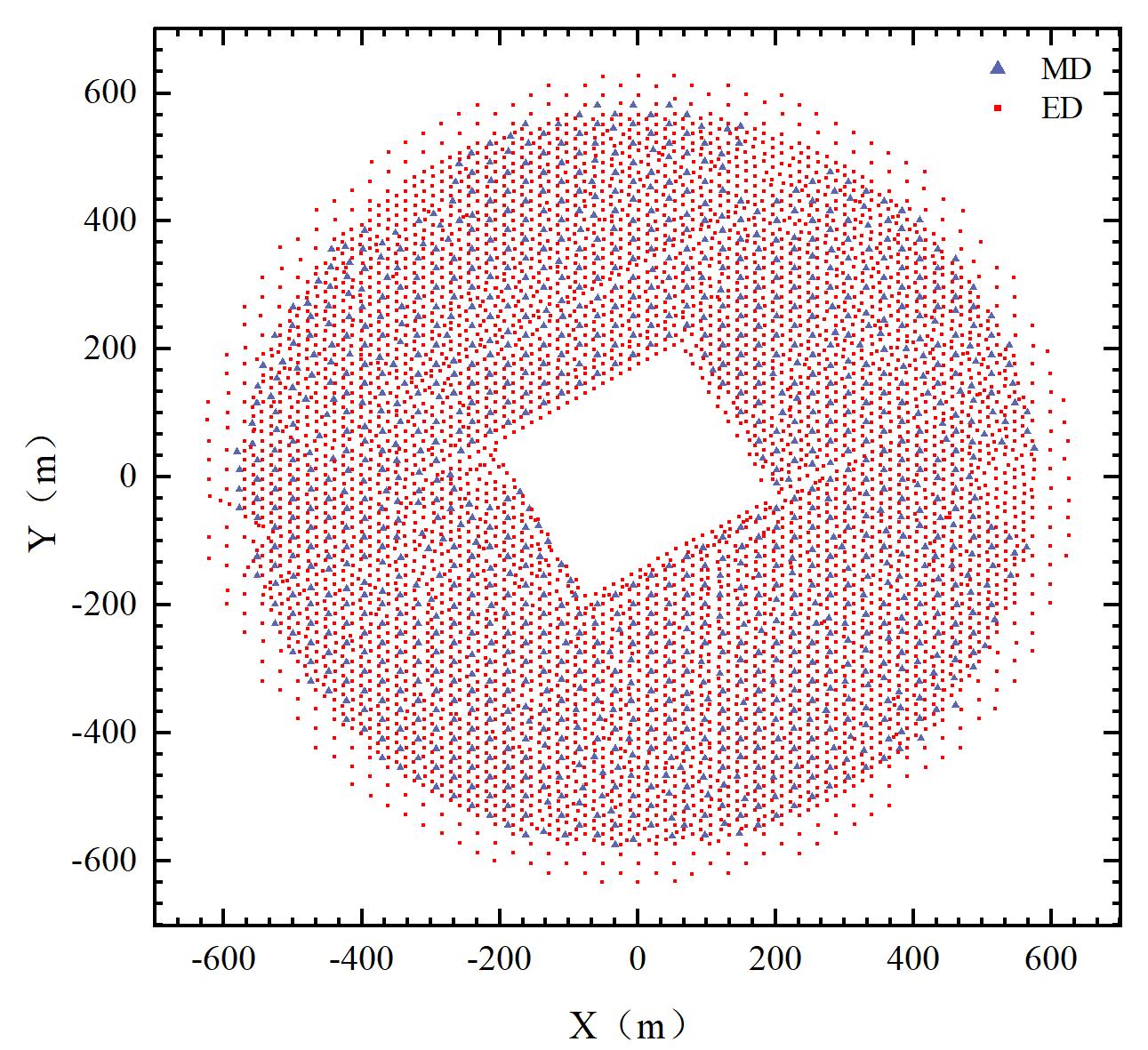} 
    \caption{\centering Layout of the whole KM2A array. The red square dots and blue triangular points indicate the EDs and MDs, respectively.}
    \label{fig1} 
\end{figure}

  The interaction of secondary particles in the detector is simulated by G4KM2A \cite{ref47}, which is developed in the framework of the GEANT4 package\cite{ref48}. It has the advantage of changing the array distribution conveniently for diﬀerent requirements. Due to the trigger rate of KM2A mainly (~99\%) determined by EDs\cite{ref49}, the responses of the ED array are simulated. Thus, an air shower event is only recorded in the simulation when more than 20 EDs are ﬁred within a time window of 400 ns, and the array is recognized to be triggered\cite{ref50}. To approximate the real conditions of KM2A, the shower core position is randomly sampled within a radius of 1000 m and the background noise of each ED is set at 1700 Hz\cite{ref51}.

\section{Simulation results and discussions}
Due to the acceleration/deceleration and deflection mechanisms of secondary charged particles caused by AEF, the number of shower events satisfying the LHAASO-KM2A trigger conditions will also change. Given these intricate correlations, comprehensive simulation studies are explored in this paper. The specific results are shown as follows.

\subsection{The shower rate variations in different near-earth thunderstorm fields}
Assuming the AEF layer with a typical thickness of 1000 m\cite{ref5}, Fig. 2 illustrates the variations of the shower rates triggered by the whole KM2A array in different electric fields. It can be seen that the shower rate increases and the enhanced amplitude becomes larger with field intensity in negative electric field. In a field of -200 V/cm, the maximum enhancement is only 1\%. However, the amplitudes could reach 5\% at -600 V/cm and 12\% at -1000 V/cm. In positive fields greater than 400 V/cm, it has been observed that the amplitudes are 
lower compared to that in negative electric fields, being 1\% in +600 V/cm and 7\% in +1000 V/cm. Additionally, in the range of 0-400 V/cm, the shower rates show a slight decrease, with the maximum amplitude being only 0.5\%.

\begin{figure}[H] 
    \centering
    \includegraphics[width=0.7\textwidth]{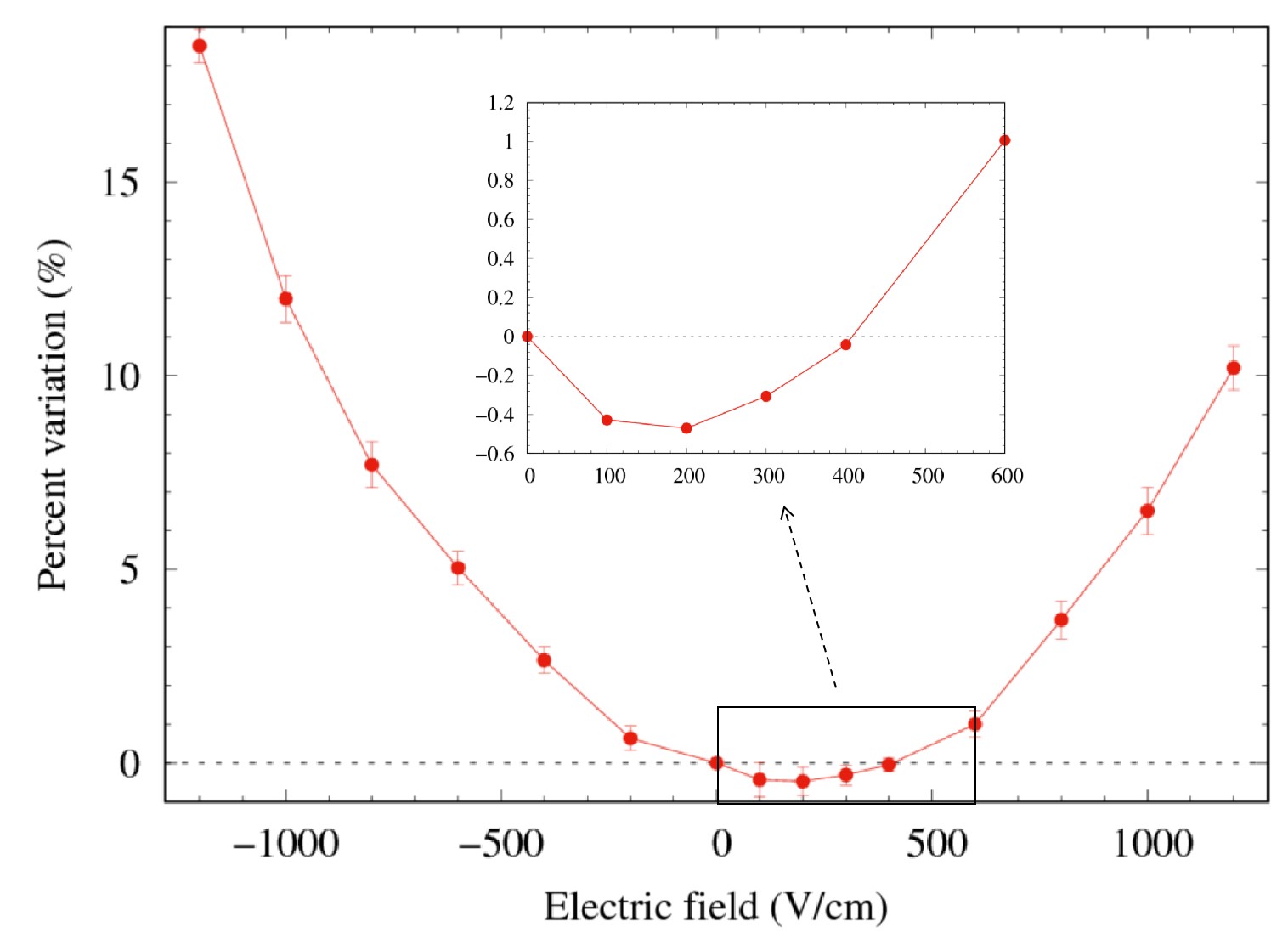} 
    \caption{ Percent variations of the shower rates as a function of the electric field.}
    \label{fig2} 
\end{figure}

To understand the change in the shower rates triggered by KM2A with the near-earth electric field, the percent changes of the average number of positrons, electrons and the sum of both in different electric fields at LHAASO are simulated. As shown in Fig. 3, when the electric field is negative (accelerating electrons), the number of electrons increases, while the number of positrons decreases. The total number of positrons and electrons increases with the electric field strength, and the amplitude enhancement is up to 32\% in an electric field of –1000 V/cm. When the field is positive (accelerating positrons), the number of electrons decreases, while the number of positrons increases. In positive fields greater than 400 V/cm, the total number increases, and the amplitude enhancement reaches 12\% in +1000 V/cm. In the range of 0−400 V/cm, the total number declines and the maximum amplitude is about 2\%. From above results, it can be concluded that if a shower event generates an equal number of electrons and positrons, and the total count will increase even if they both increase or decrease proportionally. In the same AEF, the proportion of accelerated particles is significantly higher than that of decelerated particles. Regardless of the AEF polarity, the total number of electrons and positrons tends to increase, leading to an increase of the shower rate. From Yan, Zhou et al.\cite{ref31,ref32,ref35}, we know that the number of electrons is significantly greater than that of positrons in the absence of an electric field.  Therefore, in 0-400 V/cm, the total number of positrons and electrons decreases, resulting in a decrease of the shower rate within a certain range of positive electric fields.

\begin{figure}[H] 
    \centering
    \includegraphics[width=0.7\textwidth]{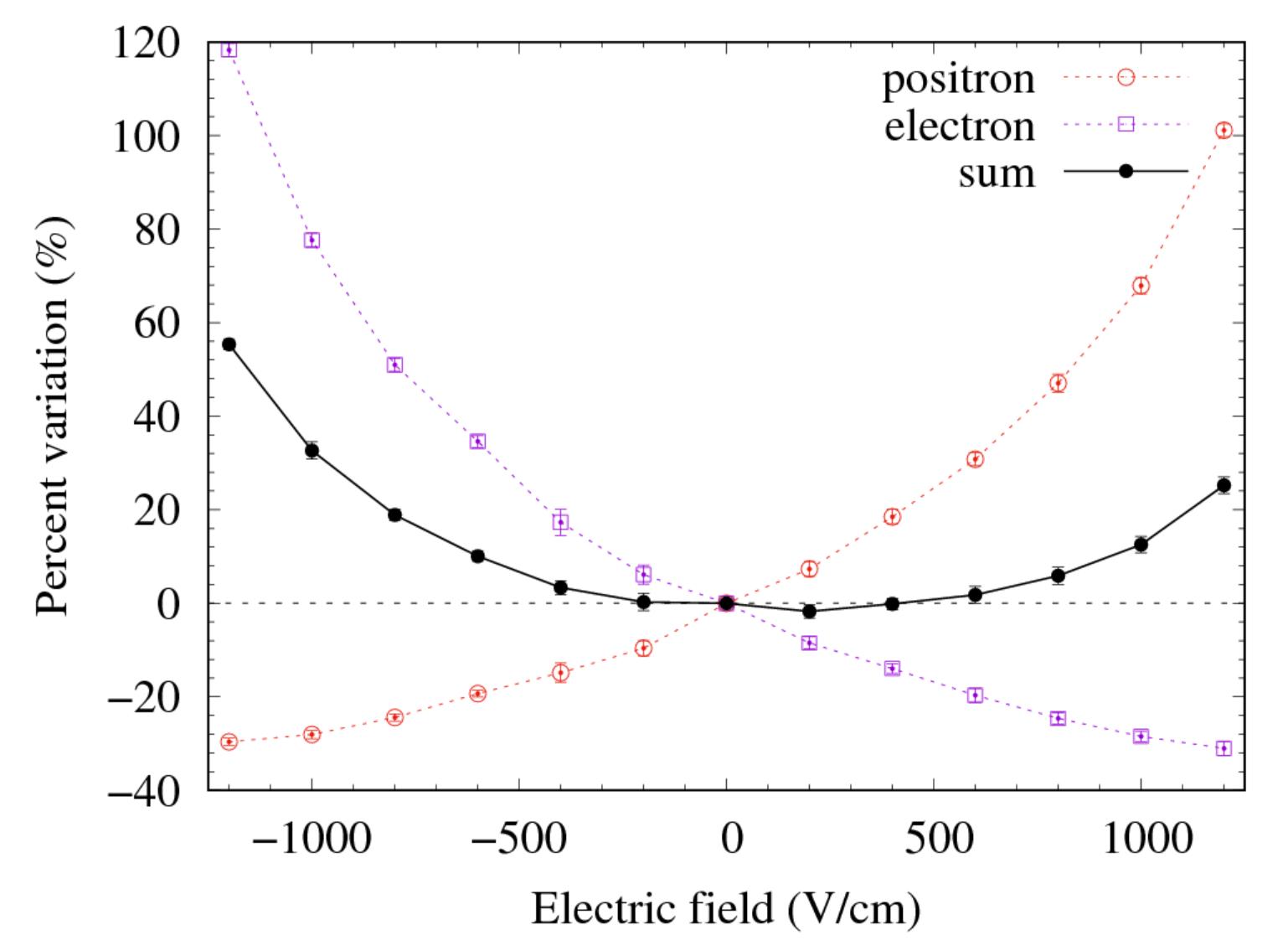} 
    \caption{\centering Percent variation of the number of positrons, electrons, and sum of both as a function of electric field, assuming an AEF layer of thickness
1000 m.}
    \label{fig3} 
\end{figure}

\subsection{The shower rate varies with AEF layer thickness}

The layer thickness of the AEF varies in different thunderstorm events and changes at various developmental stages within the same thunderstorm. Fig. 4 shows the variations of the shower rate in KM2A as a function of the vertical thickness of field layer in the atmosphere above the detector. The black circle, red triangular and blue square points correspond to the results for electric fields of -800 V/cm, -1000 V/cm, and -1200 V/cm, respectively. It can be seen that the variation behaviors of the shower rate are consistent among the three electric fields. However, the shower rate experiences a dramatic rise with thickness, and then the curve flattens out when the thickness is higher than a given value (whose value depends on the electric field strength, we define it as P$_0$). When a uniform electric field of -800 V/cm is applied, the P$_0$ is approximately 600 m, whereas P$_0$ alters to 800 m and 1000 m in an AEF of -1000 V/cm and 1200 V/cm, respectively.

\begin{figure}[H] 
    \centering
    \includegraphics[width=0.7\textwidth]{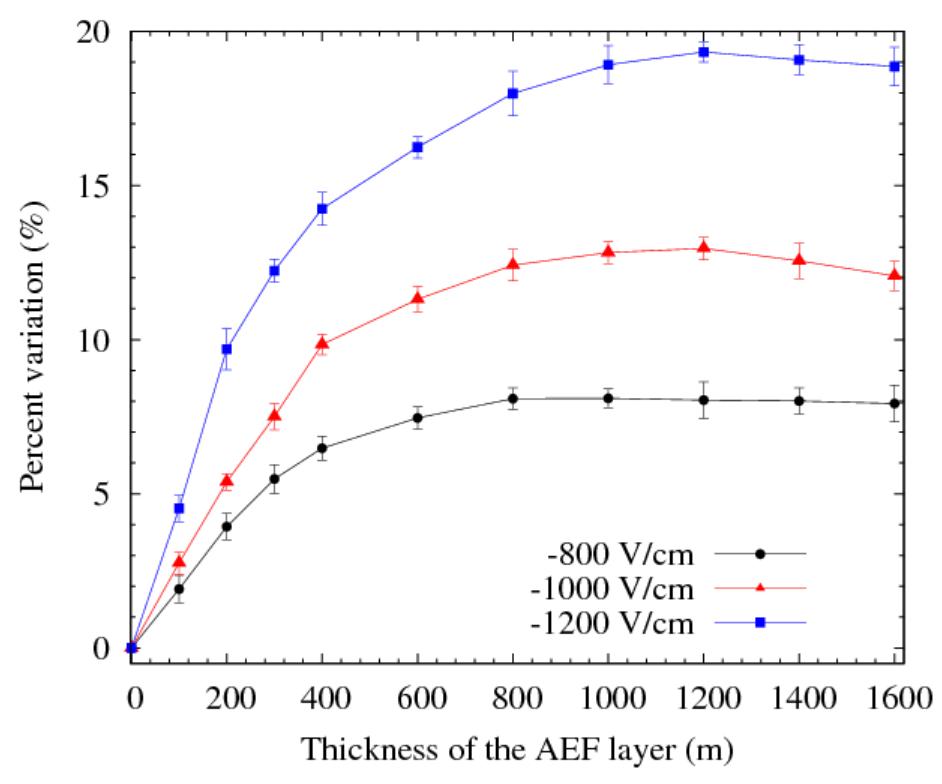} 
    \caption{Percent variations of the shower rates as a function of the thickness of AEF layer.}
    \label{fig4} 
\end{figure}

Fig. 5 shows the variations of the total number of shower positrons and electrons in -1000 V/cm, as a function of the AEF layer thickness. At small thickness, the behavior of secondary charged particles is consistent with the shower rate, showing a rapid increase. At a layer thickness of 100 m, the flux of particles rises by 7\%. and then, when the thundercloud thickness expands to 400 m, the enhanced amplitude reaches 26\%. As the thickness of the AEF layer increases, the curve gradually becomes flatten.

\begin{figure}[H] 
    \centering
    \includegraphics[width=0.7\textwidth]{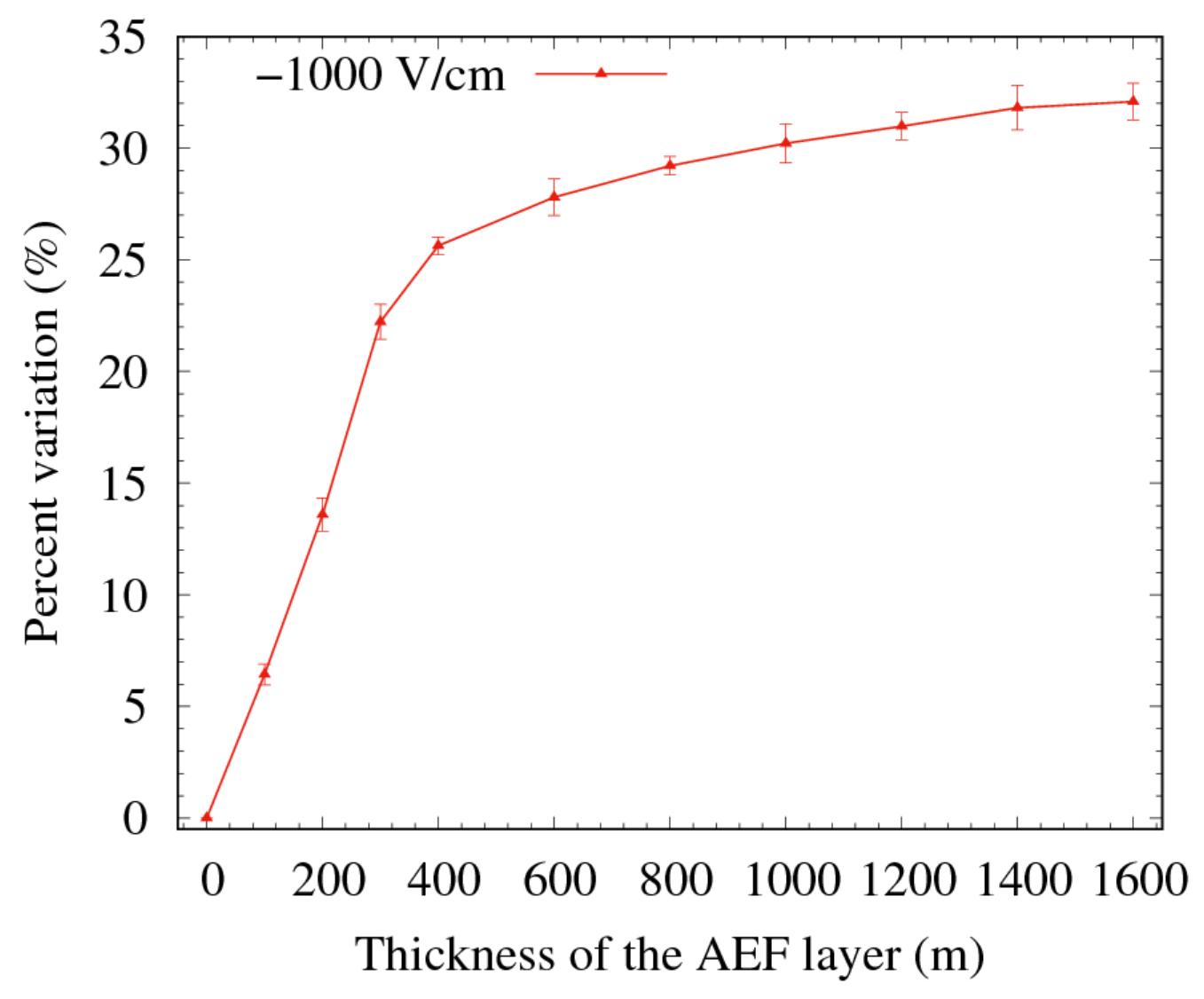} 
    \caption{ Percent variations of the number of electrons and positrons as a function of the thickness of AEF layer.}
    \label{fig5} 
\end{figure}

Fig. 6 shows the variations in the total number of positrons and electrons as a function of atmospheric depth in different AEF layer thicknesses. Firstly, we introduced a uniform atmospheric electric field of -1000 V/cm within the altitude scope of 4410 m to 6010 m (approximately 483 g/cm²). From the simulation results, we can clearly see that there is a significant increase when secondary particles enter the AEF, after developing approximately 20 g/cm², the number begins to decrease due to energy loss, but the number remains significantly higher than the number without AEF. 

\begin{figure}[H] 
    \centering
    \includegraphics[width=0.7\textwidth]{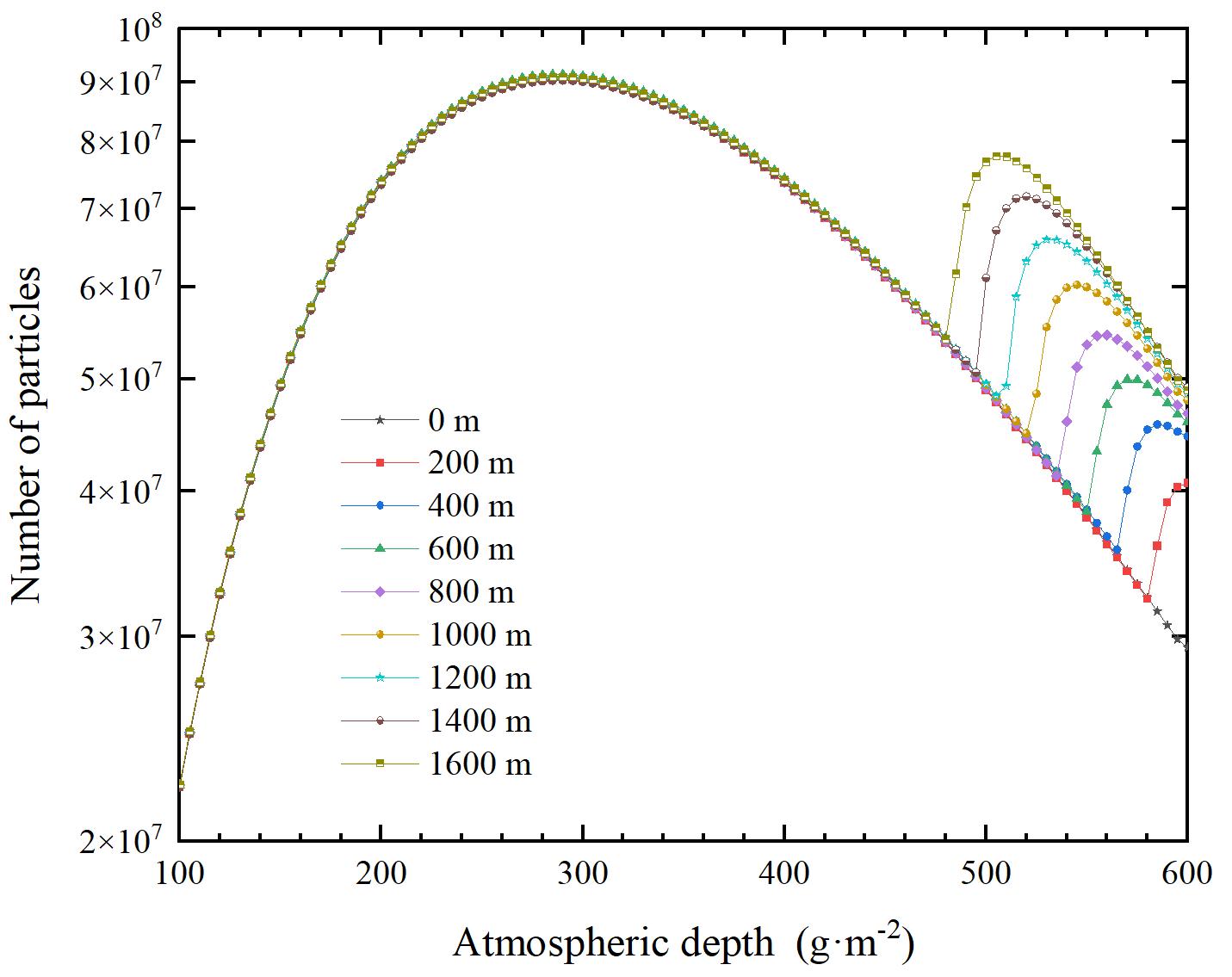} 
    \caption{ Number of electrons and positrons as a function of atmospheric depth in different layer thicknesses of AEF.}
    \label{fig6} 
\end{figure}

During thunderstorms, the secondary particles could disperse outward under the deflection of the electric field and deviate from the shower axis. In order to observe the deflection effect of the electric field on secondary particles. The CORSIKA has been used to simulate the lateral distribution of secondary positrons and electrons with various thicknesses of thundercloud layers. Fig. 7 shows the integral variations of secondary particle as a function of the distance to the shower axis ($r_{\text{max}}$) (the maximum distance of particles to the shower axis, related to the size of the detector) for different layer thicknesses in -1000 V/cm. Near the shower core, we observe that the particles decline, and the decreased amplitude becomes more significant with the layer thicknesses, the maximum amplitude reaches 17\% and 31\% when the layer thickness is 200 m and 1600 m, respectively. This implies that with a smaller detector, the particle number will notably decrease with the layer thicknesses compared to the results without an electric field, due to the deflection effect of the electric field. As the $r_{\text{max}}$ increases, the decreased amplitude gradually reduces until it disappears. When $r_{\text{max}}$ > 40 m, the number at each layer thickness begins to increase and the amplitude decreases with the layer thicknesses increase. When 110 < $r_{\text{max}}$ < 400 m, some secondary particles are deflected away from the detector because of the deflection effect of AEF, resulting in the curve of secondary particles to flatten at large layer thicknesses, corresponding to the changes of the shower rate described in Fig.4. Finally, when the detector is large enough, the shower rate will increase with the layer thickness and the curve increases slowly at large layer thicknesses.  

\begin{figure}[H] 
    \centering
    \includegraphics[width=0.7\textwidth]{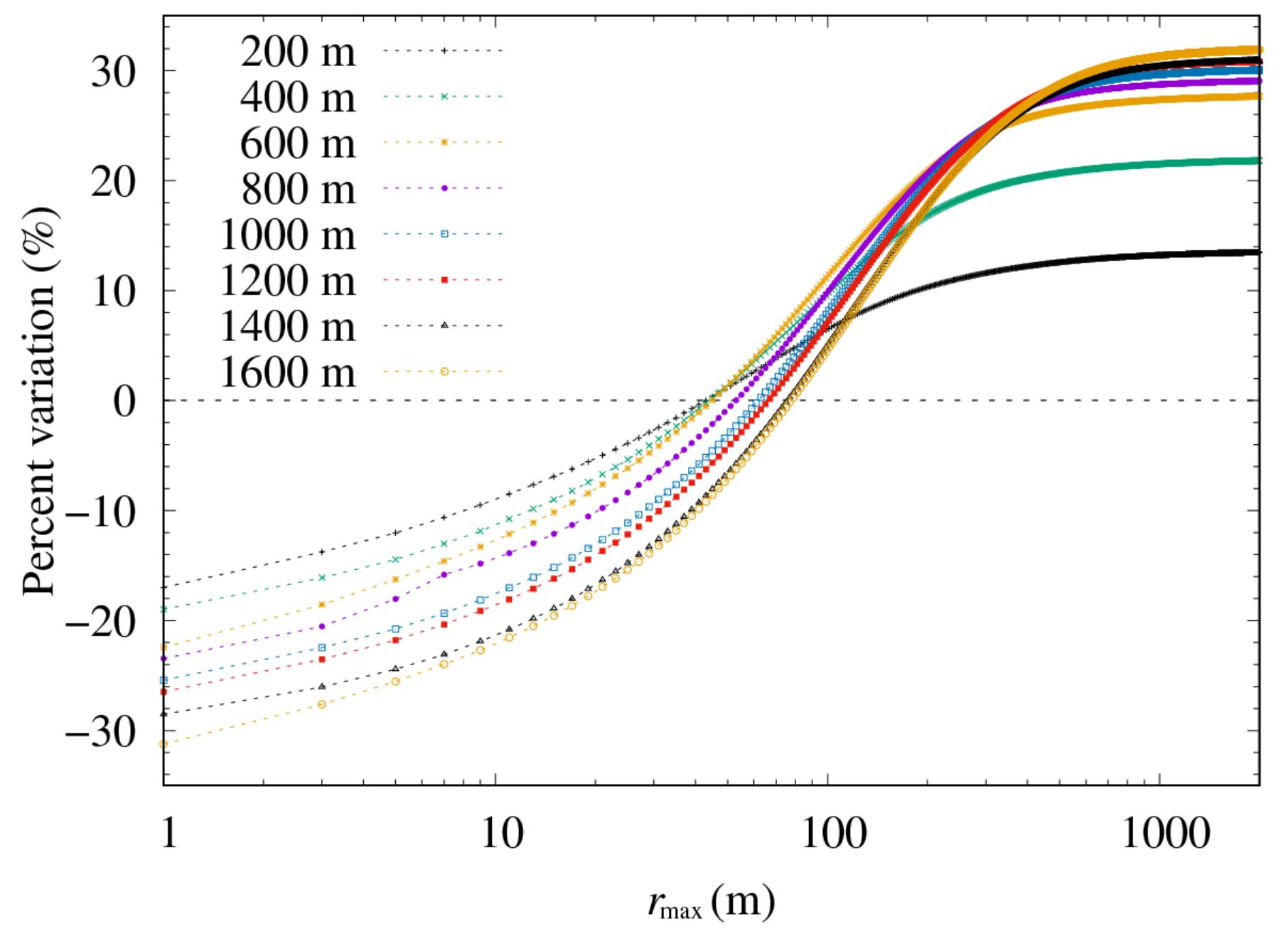} 
    \caption{\centering Percent variations of the number of electrons and positrons as a function of distance to shower axis for different layer thicknesses.}
    \label{fig7} 
\end{figure}

\subsection{ The zenith angle dependence of the shower rate variations}
It is well known that the atmospheric depth increases with the zenith angle\cite{ref36}. Therefore, the effect to which secondary cosmic rays are "absorbed" and "scattered" by the atmosphere will vary with different zenith angles. Fig. 8 shows the variations of the shower rate in KM2A as a function of zenith angle in the field of ±1000 V/cm. It can be seen that the behavior of the shower rate variation is dependent on the primary zenith angle. The rate variations of shower enhance for smaller zenith angle ranges but reduce for higher ones. In +1000 V/cm, the enhanced amplitude is up to 10\% for $\theta$ = 20°. However, the reduced behavior occurs for zenith angles larger than 40°, and the decreased amplitude is about 28\% for $\theta$ = 60°. In an electric field of -1000 V/cm, the enhancement value is 16\% at $\theta$ = 20°, larger than that in a positive field with the same strength.

\begin{figure}[H] 
    \centering
    \includegraphics[width=0.7\textwidth]{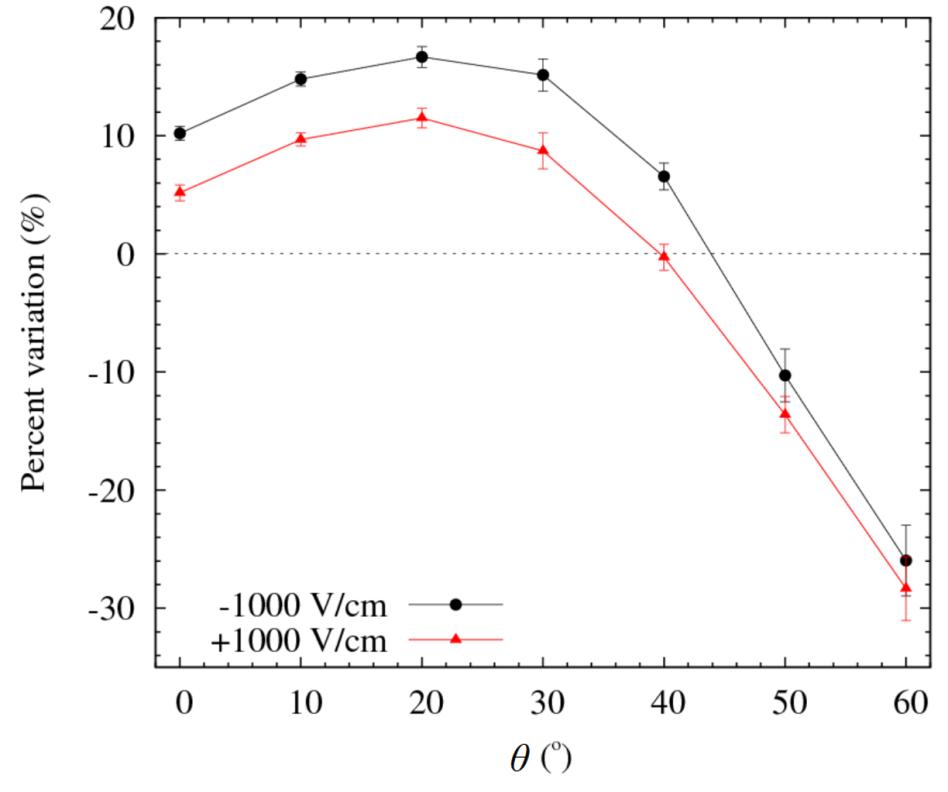} 
    \caption{\centering Percent variations of the shower rate as a function of the zenith angle, assuming an AEF layer of thickness 1000 m.}
    \label{fig8} 
\end{figure}

To understand the field effect on the shower rate with different zenith angles, we studied the field deflection effects on secondary electrons and positrons with CORSIKA. Fig. 9 (a) reports the relationship between the percent variation of particle number with different zenith angle ranges and the distance to the shower axis in the field of -1000 V/cm. It can be found that particle number decreases in lower distance to the shower axis ranges, and the decreasing magnitude amplifies with the zenith angle. The different situation occurs (which depends on the zenith angle) as $r$ increases. Until $r$ > 60 m, the number for all zenith angles increases, and the enhanced amplitude becomes larger with the smaller zenith angles. Finally, the amplitude increases with the zenith angle if the distance to the shower axis is large enough.

Fig. 9 (b) presents the integral change of Fig. 9 (a). For different $r_{\text{max}}$, the trend of the variations can be opposite. The number of particles declines for $r_{\text{max}}$ < 20 m, and the declining amplitude increases with the zenith angle. When 20 m < $r_{\text{max}}$ < 120 m, the number of secondary particles increases at smaller zenith angle ranges and decreases at larger ones, it is consistent with the variation law of the shower rate in KM2A. As the $r_{\text{max}}$ increases, the rate begins to increase, and the amplitude increases with the zenith angle decreases. Until $r_{\text{max}}$ > 450 m, the behavior reverses, and the rate increment increases with the elevation of the zenith angle. Therefore, for detectors with different sizes, the relationship between the shower rate variations and zenith angles will also differ.

From Fig. 9, we can see that the intensity variations of ground cosmic rays are highly dependent on the core distance and the primary zenith angle. In the field of -1000 V/cm, the total number of secondary particles increases for all zenith angles if the range of $r_{\text{max}}$ is not limited. Therefore, the acceleration effect in electric field is more efficient on the larger zenith angles. In the region closer to the shower core, particle numbers reduce, and the decreased amplitude increases with zenith angle. This means that the deflection effect in the field will play a major role in larger zenith angle. Hence, the changing behaviors of the shower rates in different zenith angle ranges can be understood.

\begin{figure}[H]   
\centering            
\subfloat   
{
\label{fig:subfig1}
\put(35,25){(a)}
\includegraphics[width=0.5\textwidth]{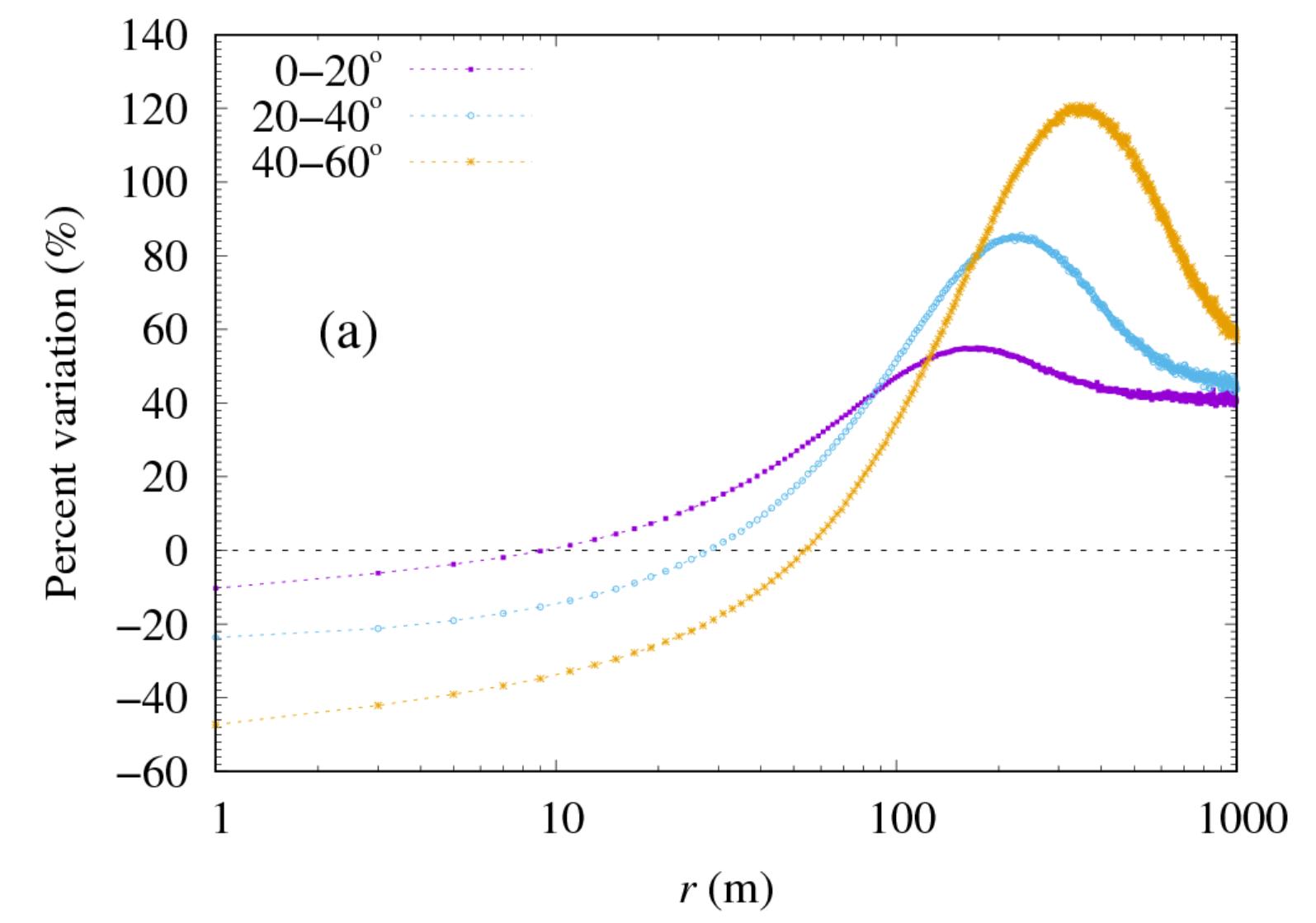}
}
\subfloat
{
\label{fig:subfig2}
\put(35,25){(b)}
\includegraphics[width=0.5\textwidth]{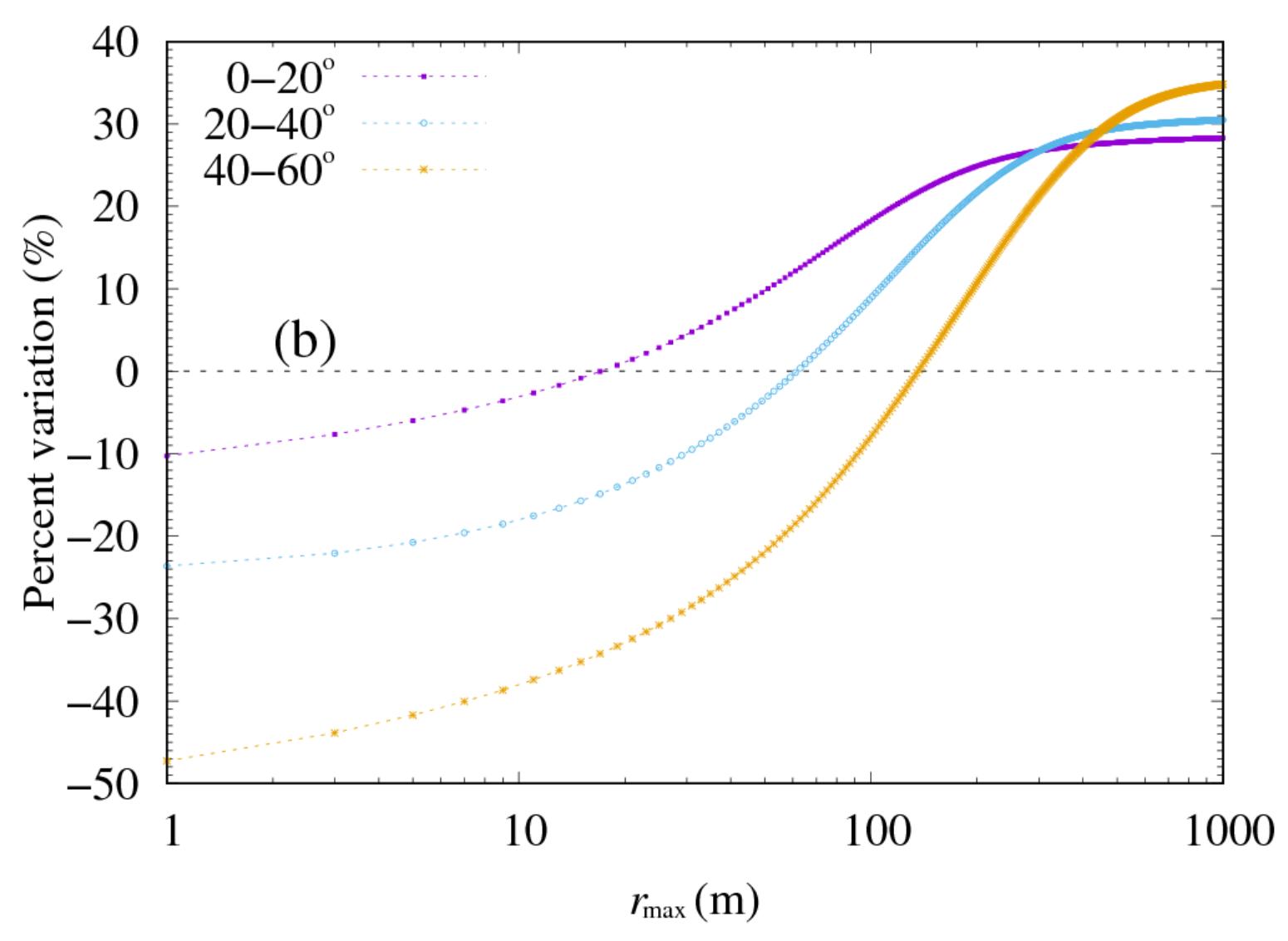}
}
\caption{\centering Percent variations of the number of electrons and positrons as a function of distance to shower axis for different zenith angle ranges (a) and corresponding integral result (b).}    
\label{fig:subfig_1}            
\end{figure}

\subsection{The primary energy dependence of the shower rate variations}
Fig. 10 shows the rate variations of shower event with different primary energies in ±1000 V/cm. It is distinctly observed that the shower rate increases significantly at lower primary energies during thunderstorms. In -1000 V/cm, the amplification of the shower rate could approximately be up to 39\% when the primary energy is 5 TeV, the enhanced amplitude only reaches 3\% for 100 TeV. The same law is observed in corresponding positive field, but the amplitude enhancement is smaller.

\begin{figure}[H] 
    \centering
    \includegraphics[width=0.62\textwidth]{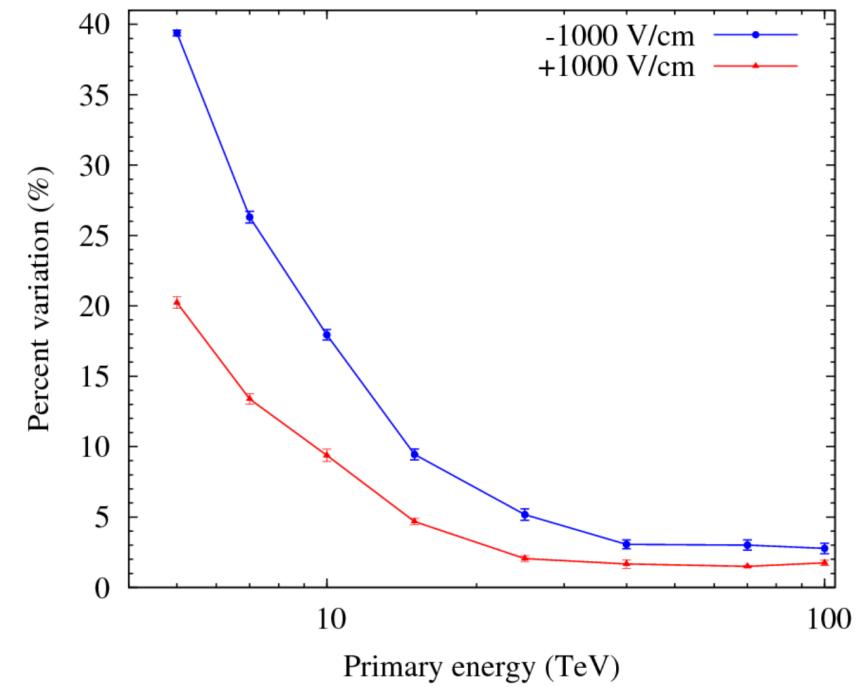} 
    \caption {\centering Percent variations of the shower rate as a function of the primary energy, assuming the AEF layer of thickness 1000 m.}
    \label{fig10} 
\end{figure}

From Fig. 10, we can see that the thunderstorm electric field has more effect on the shower event with small primary energy. Fig. 11 shows the number of shower events triggered by KM2A as a function of primary energy during thunderstorms. It can be seen that the shower events with lower primary energy show a significant increase, whereas events with higher primary energy are hardly affected during thunderstorms. As represented by the black square points in the figure, when the primary energy is 14 TeV, approximately 700 shower events are detected by KM2A during fair weather. When uniform electric field of -1000 V/cm is added, the number of detected shower events increases to about 800.

\begin{figure}[H] 
    \centering
    \includegraphics[width=0.7\textwidth]{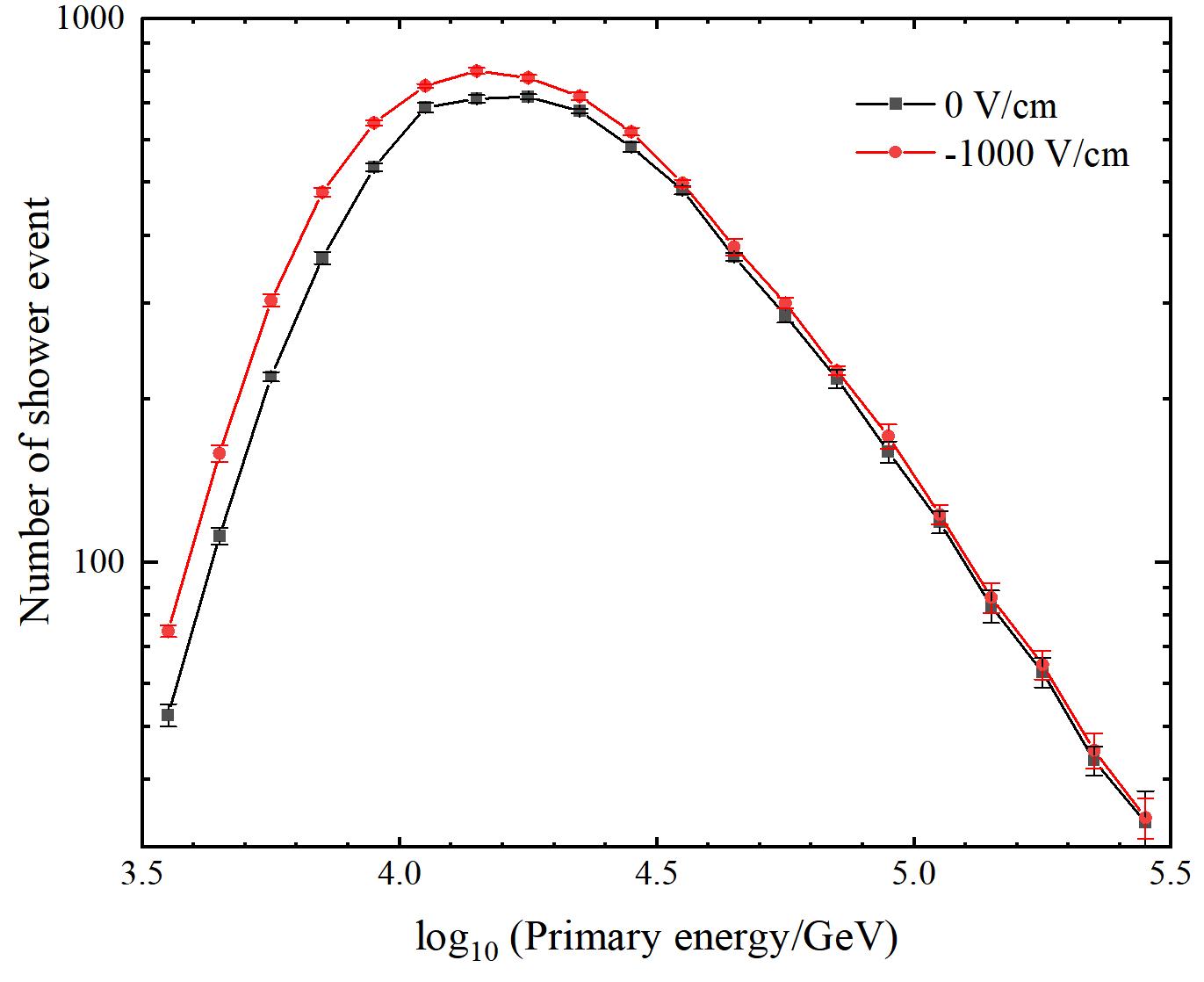} 
    \caption{\centering Energy distributions of primary proton during thunderstorms in KM2A, the results in fair weather are plotted for comparison.}
    \label{fig11} 
\end{figure}

In ground-based experiments, these low-energy particles are accelerated by the electric field, gain energy, and are successfully detected. As a result, the rate of shower events with lower primary energy has increased significantly in fields.

\section{ Conclusions}
In this study, Monte Carlo simulations have been performed with CORSIKA 7.7410 and G4KM2A packages to investigate the flux variations of cosmic ray air shower events in KM2A during thunderstorms. The simulations revealed that the variations in shower rate are intricately linked to several factors: electric field strength and polarity, layer thickness of thundercloud, primary zenith angle, and primary energy. 

During thunderstorms, the shower rate in KM2A increases, and the amplitude increases with field intensity. For identical field strengths, the enhanced amplitude is larger in negative fields. Specifically, at a field of -1000 V/cm, the increase can reach 12\%, compared to only 7\% in corresponding positive fields. In our simulation, the shower rate shows a certain degree of decrease in the positive electric field range of 0-400 V/cm. The shower rates are also influenced by the thickness of the AEF layer, we find that the trigger rate increases fast with thickness, until the thickness above a critical value (which varies with the electric field), the growth curve begins to level off. This value is about 600 m and 1000 m when the electric field strength is 800 V/cm and 1200 V/cm, respectively. This suggests that particles are significantly accelerated by the electric field within the thundercloud layer situated several hundred meters above the detector. Nonetheless, when the layer thickness of AEF is sufficiently large, the effect of the electric field on particle deflection results in no substantial increase in shower event rate detected by ground-based experiments.

Furthermore, we find that the shower rate increases at smaller zenith angle ranges, and decreases at larger ranges. By using CORSIKA, we observe that the number of secondary particle near the shower core notably reduces as the zenith angle increases, due to the deflection effect of electric field. Then, as the distances to shower core increase, the rates decrease in larger zenith angles and increase in smaller zenith angles. Finally, when the distance is large enough, the rate increases, and the enhanced amplitude increases with the zenith angle.

Meanwhile, we find that the shower rate quickly increases in lower primary energies and the enhanced amplitude of the shower rate decreases with primary energy. In -1000 V/cm, the shower event rate increases by up to 39\% when primary energy is 5 TeV, however, the enhanced amplitude dropped to 3\% for 100 TeV. We believe that strong electric field has a greater impact on lower-energy secondary particles from lower primary cosmic rays. As a result, the shower rate with lower primary energy has increased significantly during thunderstorms.

Our simulation results in this work give more information about the acceleration and deflection mechanism of secondary charged particles caused by an atmospheric electric field. Furthermore, these results offer valuable insights for understanding the shower rate variations detected by LHAASO-KM2A during thunderstorms.

\section*{Acknowledgement}
This work is supported by the National Natural Science Foundation of China
(NSFC No. 12375102 and No. U2031101).


\begin{thebibliography}{99}


\bibitem{ref1}
D. X. Liu et al., Characteristics of lightning radiation source distribution and charge structure of squall line, Acta Phys. Sin. 62, 219201 (2013).

\bibitem{ref2}
M. Stolzenburg et al., Electric field values observed near lightning flash initiations, Geophys. Res. Lett. 34, L04804 (2007).
\bibitem{ref3}
T. C. Marshal et al., Observed electric fields associated with lightning initiation, Geophys. Res. Lett. 32, L03813 (2005).
\bibitem{ref4}
R. C. T. Wilson, The electric field of a thundercloud and some of its effects, Proc. Phys. Soc. London 37, 32D (1924).
\bibitem{ref5}
A. V. Gurevich et al., Runaway electron mechanism of air breakdown and preconditioning during a thunderstorm, Phys. Lett. A 165, 463 (1992).
\bibitem{ref6}
L. P. Babich et al., New data on space and time scales of relativistic runaway electron avalanche for thunderstorm environment: Monte Carlo calculations, Phys. Lett. A 245, 460 (1998).
\bibitem{ref7}
J. R. Dwyer et al., High-energy atmospheric physics: Terrestrial gamma-ray flashes and related phenomena, Space Sci. Rev. 173, 133 (2012).
\bibitem{ref8}
J. R. Dwyer, A fundamental limit on electric fields in air, Geophys. Res. Lett. 30, 2055 (2003).
\bibitem{ref9}
E. M. D. Symbalisty et al., Finite volume solution of the relativistic Boltzmann equation for electron avalanche studies, IEEE Trans. Plasma Sci. 26, 1575 (1998).
\bibitem{ref10}
E. A. Yuzhakova et al., Lateral Distribution Functions of the Electron–Photon Component of the EAS Registered with the NEVOD-EAS Array, Physics of Atomic Nuclei 86, 1035 (2024).
\bibitem{ref11}
Axikegu et al., Cosmic ray shower rate variations detected by the ARGO-YBJ experiment during thunderstorms, Phys. Rev. D 106, 022008 (2022).
\bibitem{ref12}
G. J. Fishman et al. Discovery of intense gamma-ray flashes of atmospheric origin, Science 264, 1313 (1994).
\bibitem{ref13}
N. S. Khaerdinov et al., Electric field of thunderclouds and cosmic rays: evidence for acceleration of particles (runaway electrons), Atmospheric Research 76, 346 (2005).
\bibitem{ref14}
F. Fuschino et al., High spatial resolution correlation of AGILE TGFs and global lightning activity above the equatorial belt, Geophys. Res. Lett. 38, L14806 (2011).
\bibitem{ref15}
M. Marisaldi et al., Detection of terrestrial gamma ray flashes up to 40 MeV by the AGILE satellite, J. Geophys. Res. 115, A00E13 (2010).
\bibitem{ref16}
T. Neubert et al., A terrestrial gamma-ray flash and ionospheric ultraviolet emissions powered by lightning, Science 367, 183 (2020).
\bibitem{ref17}
T. Neubert et al., The ASIM mission on the international space station, Space Sci. Rev. 215, 26 (2019).
\bibitem{ref18}
M. S. Briggs et al., First results on terrestrial gamma ray flashes from the fermi gamma-ray burst monitor, J. Geophys. Res. 115, A07323 (2010).
\bibitem{ref19}
S. Foley et al., Pulse properties of terrestrial gamma-ray flashes detected by the fermi gamma-ray burst monitor, J. Geophys. Res. Space Phys. 119, 5931–5942 (2014). 
\bibitem{ref20}
G. J. Fishman et al., Temporal properties of the terrestrial gamma-ray flashes from the gamma-ray burst monitor on the fermi observatory, J. Geophys. Res. Space Phys. 116, A07304 (2011). 
\bibitem{ref21}
V. V. Bogomolov et al., Observation of Terrestrial gamma ray flashes in the RELEC space experiment on the Vernov satellite. Cosmic Res. 55, 159–168 (2017).
\bibitem{ref22}
H. Zhang et al., On the terrestrial gamma-ray flashes preceding narrow bipolar events Geophys Res. Lett. 48, 8 (2021).
\bibitem{ref23}
M. Marisaldi et al., Properties of terrestrial gamma ray flashes detected by AGILE MCAL below 30 MeV. Journal of Geophysical Research-Space Physics, 119, 1337–1355 (2014).
\bibitem{ref24}
V. V. Alexeenko et al., Transient variations of secondary cosmic rays due to atmospheric electric field and evidence for prelightning particle acceleration, Phys. Lett. A 301, 299 (2002).
\bibitem{ref25}
S. Vernetto, The EAS counting rate during thunderstorms, in Proceedings of 27th ICRC Vol. 10, p. 4165 (2001).
\bibitem{ref26}
H. Tsuchiya et al., Observation of thundercloud-related gamma rays and neutrons in Tibet, Phys. Rev. D 85, 092006 (2012).
\bibitem{ref27}
A. Chilingarian, Thunderstorm ground enhancements– model and relation to lighting flashes, J. Atmos. Solar Terr. Phys. 107, 68 (2014).
\bibitem{ref28}
A. Chilingarian et al., Extensive air showers, lightning, and thunderstorm ground enhancements, Astropart. Phys. 82, 21 (2016).
\bibitem{ref29}
A. Chilingarian et al., Thunderstorm ground enhancements: multivariate analysis of 12 years of observations, Phys. Rev. D 106, 082004 (2022). 
\bibitem{ref30}
T. Toriiet al., Gradual increase of energetic radiation associated with thunderstorm activity at the top of Mt. Fuji, Geophys. Res. Lett. 36, L13804 (2009).
\bibitem{ref31}
B. Bartoli et al., Observation of the thunderstorm-related ground cosmic ray flux variations by ARGO-YBJ, Phys. Rev. D 97, 042001 (2018).
\bibitem{ref32}
X. X. Zhou et al., Effect of near-earth Thunderstorms electric field on the intensity of ground cosmic ray Positrons/electrons in Tibet, Astropart.
Phys. 84, 107 (2016).
\bibitem{ref33}
F. Aharonian et al., Flux Variations of Cosmic Ray Air Showers Detected by LHAASO-KM2A During a Thunderstorm on 10 June 2021, Chin. Phys. C 47, 015001 (2023).
\bibitem{ref34}
A. Chilingarian et al., On the origin of the particle fluxes from the thunderclouds: Energy spectra analysis, Europhys. Lett. 106, 59001 (2014).
\bibitem{ref35}
R. R. Yan et al., Effects of thunderstorms electric field on energy of cosmic rays at LHAASO, Chin. Astron. Astrophys. 44, 146 (2020).
\bibitem{ref36}
K. G. Axi et al., Intensity variations of showers with different zenith angle ranges during thunderstorms, Astrophys. Space Sci. 367, 30 (2022).
\bibitem{ref37}
S. Buitink et al., Monte Carlo simulations of air showers in atmospheric electric fields, Astropart. Phys. 33, 1-12 (2010).
\bibitem{ref38}
E. S. Cramer et al., An analytical approach for calculating energy spectra of relativistic runaway electron avalanches in air, J. Geophys. Res. Space Phys. 119, 7794 (2014).
\bibitem{ref39}
A. Chilingarian et al., Recovering of the energy spectra of electrons and gamma rays coming from the thunderclouds, Atmos. Res. 114, 1-16 (2012).
\bibitem{ref40}
D. Heck et al., CORSIKA: A Monte Carlo code to simulate extensive air showers, Report FZKA 6019 (1998).
\bibitem{ref41}
X. H. Ma et al., Chapter 1 LHAASO Instruments and Detector technology. Chin. Phys. C 46, 030001 (2022).
\bibitem{ref42}
F. Aharonian et al., Observation of the Crab Nebula with LHAASO-KM2A—a performance study, Chin. Phys. C 45, 025002 (2021).
\bibitem{ref43}
https://www.ngdc.noaa.gov/geomag/calculators/magcalc.shtml\#igrfwmm, retrieved 23th April 2024 
\bibitem{ref44}
X. X. Zhou et al., Effects of Thunderstorms electric field on intensity of cosmic ray electrons, Chin. J. Space Sci. 36, 49 (2016).
\bibitem{ref45}
H. Y. Zhang et al., Approaches to composition independent energy reconstruction of cosmic rays based on the LHAASO-KM2A detector, Phys. Rev. D 106, 123028 (2022).
\bibitem{ref46}
Z. Cao et al., Measurements of All-Particle Energy Spectrum and Mean Logarithmic Mass of Cosmic Rays from 0.3 to 30 PeV with LHAASO-KM2A, Phys. Rev. L 132, 131002 (2024).
\bibitem{ref47}
https://pos.sissa.it/358/219/pdf, retrieved 12 June 2024.
\bibitem{ref48}
S. Agostinelli et al., GEANT4—A simulation toolkit, Nucl. Instrum. Meth. A, 506, 250 (2003). 
\bibitem{ref49}
X. P. Zhang et al., Effects of environmental factors on the LHAASO-ED array, in Proceedings of 38th ICRC, 444, 472 (2023).
\bibitem{ref50}
Z. Li et al., Comparison of the Measurement and Simulation with KM2A Prototype Array, EPJ Web of Conferences, 208, 14006 (2019).
\bibitem{ref51}
Zhen Cao et al., Data quality control system and long-term performance monitor of LHAASO-KM2A. Astroparticle Physics 164, 103029 (2025). 




\end{thebibliography}
\end{document}